\documentclass[twocolumn,showpacs,preprintnumbers,amsmath,amssymb]{revtex4}

\usepackage{graphicx}
\usepackage{dcolumn}
\usepackage{bm}

\begin{document}

\preprint{APS/123-QED}

\title{3D particle-in-cell simulation of electron acceleration by Langmuir waves in an inhomogeneous plasma}

\author{R. Pechhacker}
\author{D. Tsiklauri}%
\affiliation{%
School of Physics and Astronomy, Queen Mary University of London, London E1 4NS, United Kingdom
}%

\date{\today}

\begin{abstract}
A possible solution to the unexplained high intensity hard x-ray (HXR) emission observable during solar flares was investigated via 3D fully relativistic, electromagnetic particle-in-cell (PIC) simulations with realistic ion to electron mass ratio. A beam of accelerated electrons was injected into a magnetised, Maxwellian, homogeneous and inhomogeneous background plasma. The electron distribution function was unstable to the beam-plasma instability and was shown to generate Langmuir waves, while relaxing to plateau formation. In order to estimate the role of the background density gradient on an unbound (infinite spatial extent) beam, three different scenarios were investigated: a) a uniform density background; b) a weak density gradient, $n_{e,R}/n_{e,L}=3$; c) a strong gradient case, $n_{e,R}/n_{e,L}=10$, where $n_{e,R}$ and $n_{e,L}$ denote background electron densities on the left and right edges of the simulation box respectively. The strong gradient case produced the largest fraction of electrons beyond $15 v_{th}$. Further, two cases (uniform and strong gradient background) with spatially localized beam injections were performed aiming to show drifts of the generated Langmuir wave wavenumbers, as suggested in previous studies. For the strong gradient case, the Langmuir wave power is shown to drift to smaller wavenumbers, as found in previous quasi-linear simulations.

\end{abstract}

\maketitle
%
\section{\label{sec:intro}Introduction}
%
The so-called 'number problem' in the context of solar flares refers to the high number of accelerated electrons necessary in order to explain spectral observations of hard X-ray (HXR) radiation from the solar corona \cite{2002SSRv..101....1A,2003AdSpR..32.1001L}. For a number density of $n=10^{16}$ m$^{-3}$ and a solar flare particle acceleration volume of $\approx 1-10$ Mm$^3$, the acceleration mechanism must be operating at $100\%$ efficiency. No such mechanism is known. A number of theories have been put forward in an attempt to solve this problem: i) re-acceleration of already slowed down electrons in the chromosphere \cite{2009A&A...508..993B}. However, observations show that a large part of the accelerated electrons drifts towards the coronal loops rather than the chromosphere \cite{2007ApJ...666.1256B}; ii) formation of an electric circuit of precipitating and returning electrons \cite{2011SSRv..159..357Z,2011Ge&Ae..51.1029Z}; iii) dispersive Alfven waves propagating towards loop foot points and accelerating particles in plasmas with transverse density inhomogeneities \cite{2005A&A...435.1105T,2008ApJ...675.1645F,2011PhPl...18i2903T,2012PhPl...19h2903T}; iv) acceleration by Langmuir waves in non-uniform plasmas, as the Langmuir spectrum drifts to smaller wave-numbers \cite{2012A&A...539A..43K}.\\
Ref.\cite{2012A&A...539A..43K} presents a collisional quasi-linear theory study, which confirms that in the case of an inhomogeneous plasma, the generated Langmuir waves show a drift in k-space, which results in an increased number of electrons carrying higher energies. It investigates the interplay between particle collisions as well as Langmuir wave generation and absorption. Qualitatively, the main argument states that a Langmuir wave, that has been generated by an electron of a given energy, will have a wavenumber $k$ corresponding to that electron's energy. A positive density gradient in the direction of wave propagation will cause a decrease of the wavenumber, and hence higher phase velocity $v_{ph}=\omega/k$. The $k$-shifted wave is then subject to absorption by a faster electron. The overall effect is an increased number of high energy electrons in the energy spectrum. While the quasi-linear approach allows simulations for long time scales, fully kinetically self-consistent dynamics of the phase space distribution function of the electron population remains inaccessible. Such dynamics is of high relevance to the beam-plasma instability, which is claimed to be responsible for the generation of Langmuir waves. Ref.\cite{2012A&A...544A.148K} carried out three-dimensional particle-in-cell (PIC) simulations of a mono-energetic electron beam being injected into plasma. The study successfully showed the characteristic plateau formation, along with noticeable acceleration of electrons. It suggests that the total amount of energy stored in electrons moving faster than the initial beam electrons could be (depending on simulational parameters e.g. magnetic field strength) of the order of $10-30\%$. However, Ref.\cite{2012A&A...544A.148K} did not take density gradients or collisions into account, which may well alter the obtained results.\\ 
In this paper, fully relativistic, electromagnetic, collisionless 3D particle-in-cell simulations of an electron beam being injected into a Maxwellian, magnetised, non-uniform plasma are performed. This work extends Ref.\cite{2012A&A...539A..43K} by the inclusion of self-consistent plasma kinetics, while it lacks the effect of collisions. Ref.\cite{2012A&A...544A.148K} is extended by the inclusion of the effect of a density gradient.\\
In section \ref{sec:simset}, the parameters of the numerical runs are discussed. Section \ref{sec:unbb} presents results from simulations of an unbound (spatially infinite extent) beam into $3$ different background plasma density profiles: i) a uniform background density; ii) a weak gradient; iii) a strong gradient. The results are then compared. In section \ref{sec:locb}, a study of a localized beam injected into a uniform background, as well as a strong gradient case is analysed. Conclusions are drawn in section \ref{sec:conclusion}.
\\
%
\section{\label{sec:simset}Simulation Setup}
%
All simulations, presented in this paper, use EPOCH, a fully electromagnetic, relativistic particle-in-cell code that was developed by the Engineering and Physical Sciences Research Council (EPSRC)-funded collaborative computational plasma physics (CCPP) consortium of UK researchers.\\
We are considering a 3D Maxwellian plasma. The background magnetic field along the $x$-direction is kept constant $B=B_x=0.003$ T$=30$ G, setting the electron gyrofrequency to $\omega_{ce}=5.28 \times 10^8$ Hz rad everywhere. The background temperature is $T=10^6$ K in each direction and isotropic, corresponding to a thermal electron velocity of $v_{th} \approx 3.9 \times 10^6$ m/s $\approx 0.013c$. Due to computational limitations, the maximum background plasma density is $n_0=10^{14}$ m$^{-3}$, giving $\omega_{pe}=5.64 \times 10^8$ Hz rad. This sets $\frac{\omega_{ce}}{\omega_{pe}}=0.935 \approx 1$. Further, the corresponding electron Debye length is $\lambda_{De}=6.9 \times 10^{-3}$ m. The simulation setup is such that $x_{max}=10000\lambda_{De}$ and $y_{max}=z_{max}=10\lambda_{De}$, while the grid size is $\lambda_{De}$.\\
A beam of accelerated electrons is injected at simulation time $t=0$. It carries a  momentum of $p_b=p_{b,x}=m_e \gamma \frac{c}{2}$ with the Lorentz factor $\gamma \approx 1.155$, while $p_y=p_z=0$. The beam is mono-energetic. The beam electrons are not replenished, i.e. there is only an injection at $t=0$ and no further electrons are being added at any other point in the simulation. As a result of the above defined quantities, at $x=0$, the plasma beta is $\beta=3.86 \times 10^{-4}$. The mass ratio used is $m_i/m_e=1836$. We will consider cases where the beam density is uniform in the plasma box, as well as cases where the beam is spatially localized. Each particle species (background electrons, ions, and beam electrons) is represented by $100$ pseudo particles per cell. The boundaries of the simulation box in the perpendicular directions ($y$ and $z$) are periodic for waves and particles alike, however, along the $x$-axis boundaries are periodic for particles, but we use open boundaries for waves. On one hand, this allows us to keep the total number of beam particles constant throughout the simulation. On the other hand, waves that were generated at the boundaries of the simulation box and travel through the boundary will not be re-introduced at the other side. This is particularly important in runs with density gradients, because, if periodic boundary conditions were being used, these waves would encounter a sudden (non-physical) change in background density at the boundary and interact via refraction. Moreover, these waves would spread out into the simulation box and interfere with the results. This does introduce loss of energy that is carried by waves that escape the simulation box, but the loss does not turn out to be significant. In order to be able to compare runs with different background plasma frequencies, we will normalize all units (where appropriate) to the above defined quantities. Time is normalized to $\omega_{pe}^{-1}$, space to $c/\omega_{pe}$, electric field to $\omega_{pe}cm_e/e$.
\\
%
\section{\label{sec:unbb}Unbound Beam Injection}
%
In this section, a beam of constant density is injected evenly distributed in the simulation box. The background density profiles are being varied in order to study the effects of density gradients on the acceleration mechanism. The density profiles can be represented by
\begin{equation}\label{eq:bg_denprof}
  n_e(x) = n_{w,s} \exp[2x/L]
\end{equation}
with $L$ and $n_{w,s}$ being chosen for each run such that the density increase from the left boundary to the right is a factor of $3$ for the {\it weak gradient} case (corresponding to $n_w$), respectively $10$ for the {\it strong gradient} case (corresponding to $n_s$). This density profile is similar to the one chosen in Ref.\cite{2012A&A...539A..43K}. Additionally, we consider a run with {\it constant background density}, $n_c$. For the constant density case, a beam to plasma ratio of $n_b/n_c=0.05$ was chosen. The parameters $n_{w,s}$ and $n_b$ were determined by making sure that $n_b$ and the average Langmuir wave growth rate, $\gamma_{LM} \propto n_b/n_e$, are equal in all runs. This ensures that the runs are comparable, as the same amount of energy is injected into the plasma in all runs, while the overall growth rate for the Langmuir waves is also equal in all runs. Naturally, in the non-zero gradient runs, local growth rates will differ from the overall rate. Furthermore, EPOCH requires the grid size to be smaller or equal to the Debye length. The grid size is set to be constant throughout the box and its value was set in the previous section. As a result, the density at the right edge of the simulation box is set to the maximum density, $10^{14}$m$^{-3}$, fulfilling the requirement everywhere. This yields $n_b=1.954 \times 10^{12}$m$^{-3}$ for all runs.
\\
%
\subsection{\label{sec:f0}Constant Background Plasma}
%
The constant background density is $n_c=3.91 \times 10^{13}$ m$^{-3}$. Injecting a uniform beam into the plasma generates a constant flow of electrons throughout the simulation box. As the background distribution is Maxwellian, the beam will introduce a return current throughout the box from the start of the simulation. This current will cause a standing wave, visible in the $E_x$-component of the electric field. The frequency of this wave is the local electron plasma frequency. It is possible to start from a zero current situation by introducing a backward drift of the background plasma to balance out the current produced from the beam electrons. However, the local backwards drift would be a function of the local beam-to-background electron ratio, $v_e = -v_b n_b / n_e$, and in the case of background density gradients, it would also cause different parts of the background plasma to drift at different velocities, which would influence the simulation results. Effectively, denser regions would be drifting slower than less dense ones, which would distort the density profile more and more over the course of the simulation, making the analysis of the role of the density gradient impossible. Therefore, we would like to keep simulations with different background density profiles as comparable as possible, changing only a minimal amount of parameters. Conveniently, it turns out that the amplitude of the wave, which is being generated by the non-zero initial current, is overpowered by the generated Langmuir waves in all runs, when given enough time for the beam-plasma instability to take effect. 
\\
%
\subsubsection{Electric Field Evolution}
%
Movie 1 in Ref.\cite{mov:ref} shows the evolution of the beam and background densities over the course of the simulation. We can see that despite the constant density ratios everywhere, wave growth occurs first from the left side of the simulation box. This is probably due to the not $100\%$ smooth periodic boundaries, which might prove just enough distortion to jump start the Langmuir wave growth.
\begin{figure}
\includegraphics[scale=0.49]{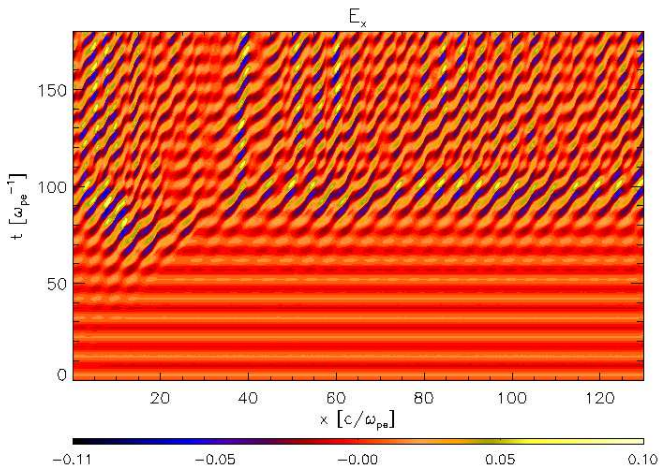}
\caption{\label{fig:tdf0} Time-distance plot for $E_x$ for constant background density and uniform beam injection along the $x$-direction; $y=z=5 \lambda_{De}$.}
\end{figure}
Fig.\ref{fig:tdf0} shows the time-distance plot for the $x$-component of the electric field along the $x$-direction; $y=z=5 \lambda_{De}$. Clearly, there is only little excitation of the electric field in the early stages of the simulation, where only effects due to the non-zero initial current take place. After $\approx 70 \omega_{pe}^{-1}$ waves are being generated due to the beam-plasma instability. The Langmuir waves can be clearly distinguished due to their propagation, opposed to the stationary nature of the waves generated by the non-zero current.
\begin{figure}
\includegraphics[scale=0.49]{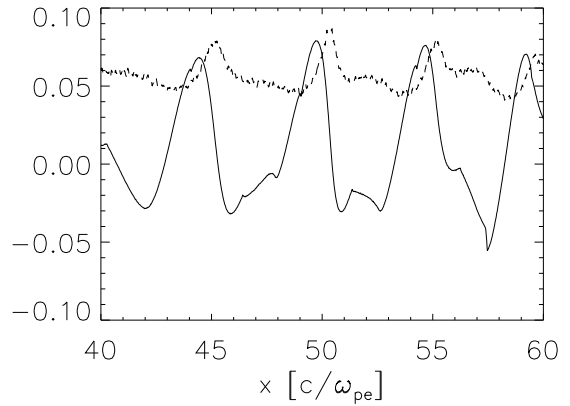}
\caption{\label{fig:exnef0} $E_x(x)$ (solid) and $n_e(x)$ (dashed) at $t=100 \omega_{pe}^{-1}$, zoomed in and arbitrary units on the $y$-axis for clarity.}
\end{figure}
Fig.\ref{fig:exnef0} compares snapshots of $E_x$ (solid) and $n_e$ (dashed) at $t=100 \omega_{pe}^{-1}$. It can be deduced that the shown waves are electrostatic, as they clearly follow Gauss' law, $\nabla \cdot {\bf E} = \rho/\epsilon_0$. Maxima and minima in the electron density profile correspond to the points of maximum gradient in the $E_x$ profile. The fact that the waves fulfil Gauss's law shows that the waves are electrostatic, i.e. Langmuir waves. 
\begin{figure}
\includegraphics[scale=0.49]{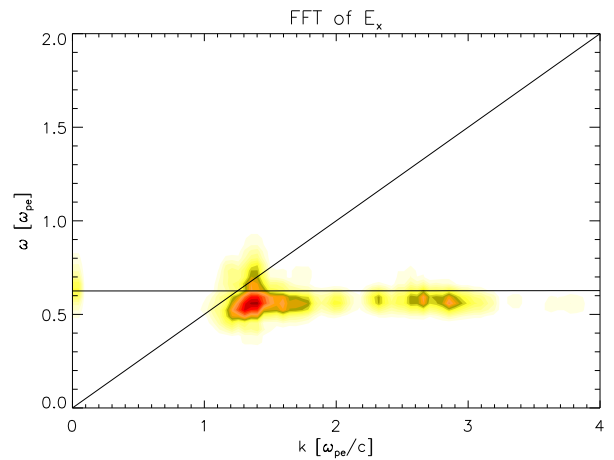}
\caption{\label{fig:fftf0} Fast-Fourier transform of $E_x$ for constant background density and uniform beam injection. The curves represent the dispersion relation for Langmuir waves, $\omega=\sqrt{\tilde{\omega}_{pe}^2 + 3v_{th}^2k^2}$ ('horizontal' curve), and the resonance condition for the beam plasma instability, $\omega = kv_b$.}
\end{figure}
Fig. \ref{fig:fftf0} shows the 2D-Fourier transform spectrum of Fig. \ref{fig:tdf0}. Predominantly, waves are being excited at the average plasma frequency, $\tilde{\omega}_{pe}$. Note that $\omega_{pe}$ always refers to the value given in the section \ref{sec:simset} and is chosen in order to be able to compare time scales in different runs, whereas $\tilde{\omega}_{pe}$ is run-specific and allows investigation of e.g. run-specific dispersion relations. The figure shows a maximum roughly at the intersection of the two curves representing the dispersion relation of Langmuir waves, $\omega=\sqrt{\tilde{\omega}_{pe}^2 + 3v_{th}^2k^2}$, and resonance condition for the beam plasma instability, $\omega = kv_b$.
\\
%
\subsubsection{Distribution Function Dynamics}
%
In this section, the dynamics of the electron velocity distribution function is discussed. According to quasi-linear theory, a distribution that is unstable to the beam-plasma instability will generate Langmuir waves, while the electron distribution function relaxes to a plateau shape. The essential requirement for wave growth is that the distribution function in phase space shows a positive slope in the forward direction with respect to the magnetic field, $\frac{\partial f(v)}{\partial v_{\parallel}} > 0$. As soon as the plateau is formed, this condition is no longer fulfilled and no more waves can be generated via this instability. 
%
%
%
\begin{figure}
\includegraphics[scale=0.49]{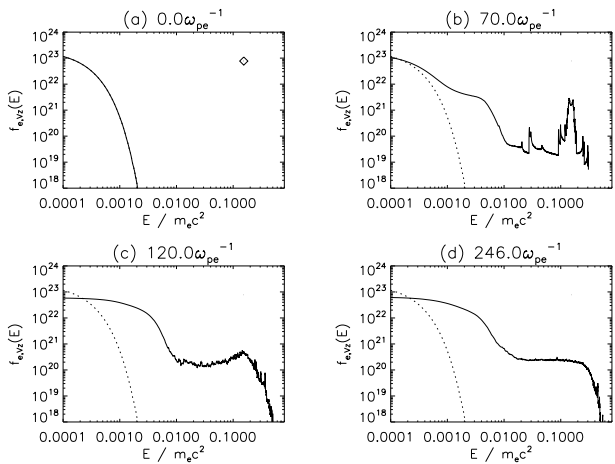}
\caption{\label{fig:especf0} Snapshots of the electron distribution function (solid) with respect to energy for constant background density and uniform beam injection. Dotted curves track the initial distribution at $t=0$.}
\end{figure}
%
The electron velocity distribution function allows us to deduce the distribution of electrons with respect to their kinetic energy, see Fig.\ref{fig:especf0}. The figure shows snapshots of $f(E_{kin},t)$ at different times in the simulation. The chosen times are representative of the initial situation, the phase of wave generation and a formed plateau at a later stage in the simulation. The initial distribution shows a Maxwellian background distribution as well as the mono-energetic beam, represented by the diamond. From Fig.\ref{fig:tdf0} we can see, that the Langmuir wave growth starts around $70 \omega_{pe}^{-1}$. The distribution function at this stage (Fig.\ref{fig:especf0}b) shows a clear deviation from the initial distribution, namely broadening of both the beam and the background distribution and a tendency towards plateau formation. At $t=120 \omega_{pe}^{-1}$, there is already a plateau formation clearly visible (Fig.\ref{fig:especf0}c). The plateau stabilizes and is still present at much later times in the simulation (Fig.\ref{fig:especf0}d). Clearly, from the point of wave generation, the amount of high energy electrons increases, which is in line with results in Ref.\cite{2012A&A...544A.148K}. Further, it can be clearly seen that some electrons are even accelerated beyond their initial injection speed and, therefore, gain energy.
\\
%
\subsection{\label{sec:f3} Weak Gradient Case}
%
According to quasi-linear theory, the background density gradient is expected to alter results significantly \cite{2012A&A...539A..43K}. Wave dissipation and refraction should be enhanced. The weak gradient case shows an increase of background plasma density from the left edge of the simulation box to the right by a factor of $3$, $n_{e,R}/n_{e,L}=3$. The shape of the density profile is defined by Eq.\ref{eq:bg_denprof}, with $n_w=2.147 \times 10^{13}$ m$^{-3}$ and $L=18204.8 \lambda_{De}$. 
\\
%
\subsubsection{Electric Field Evolution}
%
It is conceivable from movie2 in Ref.\cite{mov:ref}, that wave growth starts earliest at the left edge of the simulation box, i.e. where wave growth is most likely. Eventually, waves are being generated everywhere in the plasma.
\begin{figure}
\includegraphics[scale=0.49]{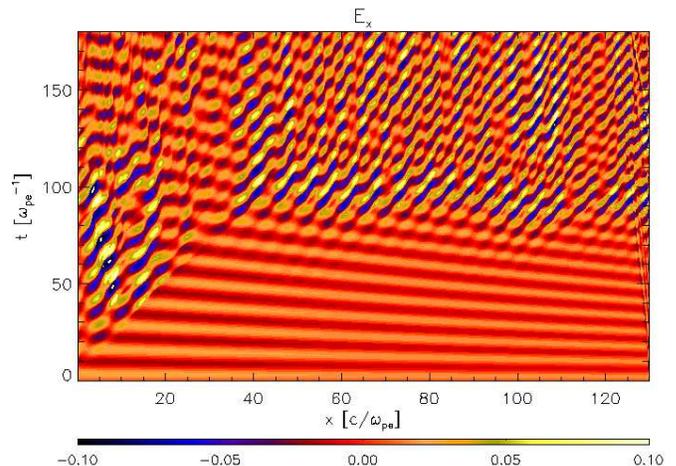}
\caption{\label{fig:tdf3} As in Fig.\ref{fig:tdf0} but for the weak gradient case.}
\end{figure}
It can be immediately seen that the initial oscillations due to the non-zero initial current is no longer uniform. However, it is still overpowered by the Langmuir wave power. The introduction of a density gradient varies the Langmuir wave growth rate accordingly. Therefore, it is to be expected that wave growth on in the left part of the simulation box (where the density is now lower) should occur earlier than in the previous section. Fig.\ref{fig:tdf3} shows the evolution of $E_x$. We can observe waves as early as $\approx 40 \omega_{pe}^{-1}$ in the less dense left part of the simulation box, while further to the right waves are being generated at a later stage in the simulation.
\begin{figure}
\includegraphics[scale=0.49]{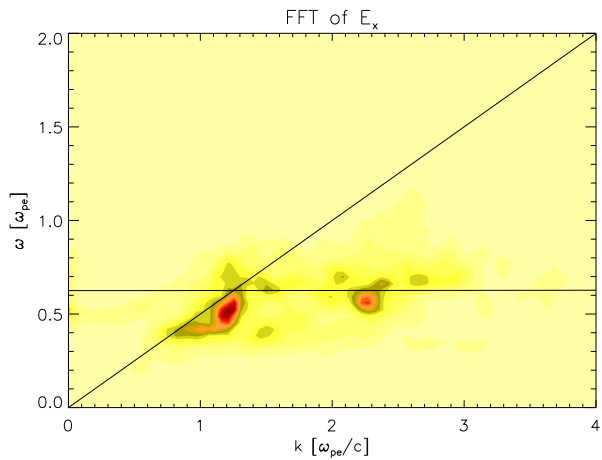}
\caption{\label{fig:fftf3} As in Fig.\ref{fig:fftf0} but for the weak gradient case.}
\end{figure}
Fig.\ref{fig:fftf3} is the Fourier transform of Fig.\ref{fig:tdf3}. It shows that wave growth is no longer as confined in ($\omega$,$k$)-space as it used to be for the constant density case. This is to be expected as the maximum growth will occur at all intersection points of the Langmuir wave dispersion relation with the beam mode. The Langmuir wave dispersion relation is a function of the local plasma frequency and therefore, also a function of density. In Fig.\ref{fig:fftf3} the 'horizontal' curve follows the dispersion relation according to the mean plasma frequency, in order to give some kind of indicator as to what frequencies are present. Different local plasma frequencies will give intersection points with the beam mode. As wave growth occurs earlier in less dense regions, the wave power is expected to be stronger in the lower frequencies, which is consistent with what can be gathered from Fig.\ref{fig:fftf3}.
\\
%
\subsubsection{Distribution Function Dynamics}
\begin{figure}
\includegraphics[scale=0.49]{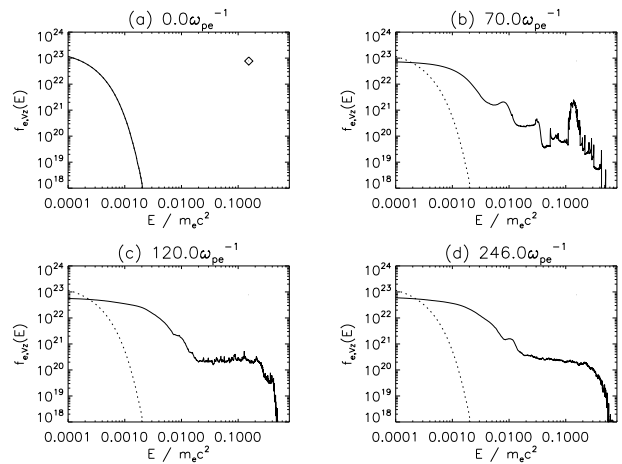}
\caption{\label{fig:especf3} As in Fig.\ref{fig:especf0} but for the weak gradient case.}
\end{figure}

%
%
Fig.\ref{fig:especf3} shows the electron distribution function for the weak gradient case. The overall distribution shows evolution similar to the constant density one. There seems to be an increased broadening of the background distribution, i.e. heating of the background plasma. 
%
\\
%
\subsection{\label{sec:f10} Strong Gradient Case}
%
The strong gradient case is defined as a $10$-fold increase of density from the left edge of the simulation box to the right, $n_{e,R}/n_{e,L}=10$. The relevant parameters for Eq.\ref{eq:bg_denprof} are $n_s=10^{13}$ m$^{-3}$ and $L=8685.89 \lambda_{De}$. Movie3 in Ref.\cite{mov:ref} shows the evolution of the densities.
\begin{figure}
\includegraphics[scale=0.49]{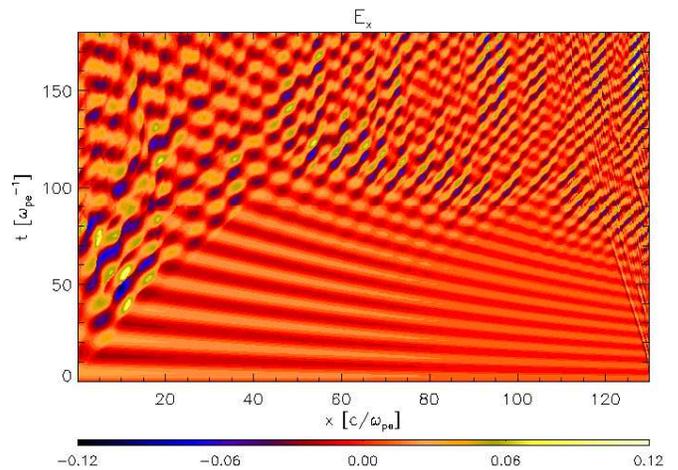}
\caption{\label{fig:tdf10} As in Fig.\ref{fig:tdf0} but for the strong gradient case.}
\end{figure}
\begin{figure}
\includegraphics[scale=0.49]{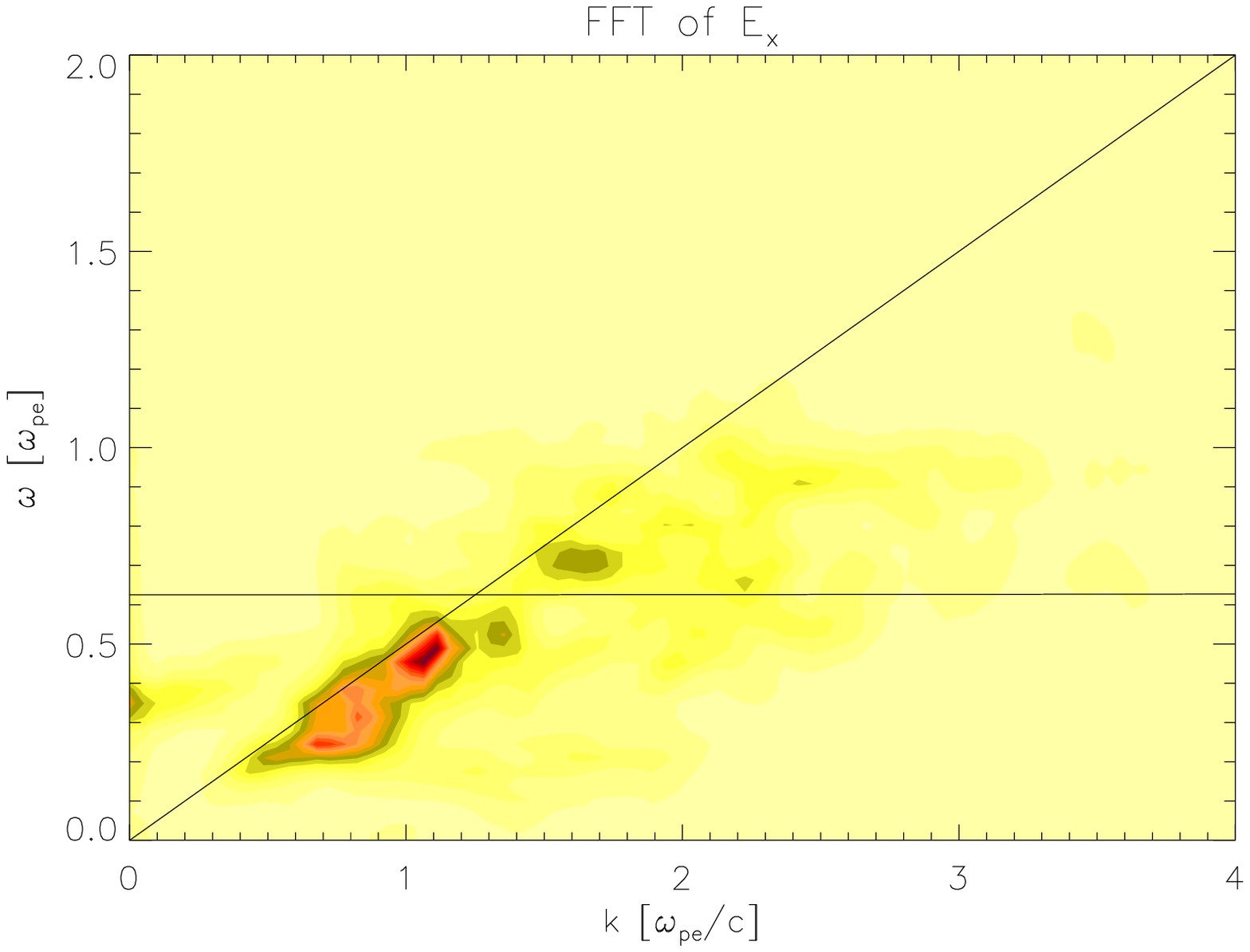}
\caption{\label{fig:fftf10} As in Fig.\ref{fig:fftf0} but for the strong gradient case.}
\end{figure}
%
%
%
%
\begin{figure}
\includegraphics[scale=0.49]{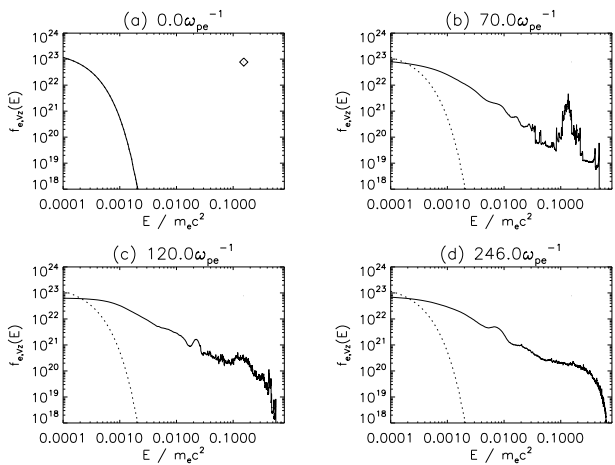}
\caption{\label{fig:especf10} As in Fig.\ref{fig:especf0} but for the strong gradient case.}
\end{figure}
Figs.\ref{fig:tdf10}-\ref{fig:especf10} show results corresponding to the strong gradient case. Fig.\ref{fig:tdf10} shows a yet more complex picture of the $E_x$-component. The increased density gradient gives  rise to a wider range of excited frequencies, see Fig.\ref{fig:fftf10}. General behaviour of the distribution functions is along the lines of the weak gradient, but heating is yet more pronounced. Additionally, the fraction of accelerated electrons is larger, which is shown more evidently in the following section.
\\
%
\subsection{Comparison}
%
\begin{figure}
\includegraphics[scale=0.49]{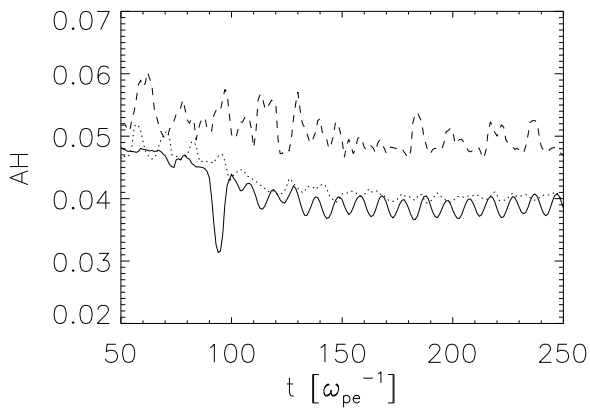}
\caption{\label{fig:apf} Fraction of electrons with energies higher than $(15 v_{th})^2 m_e \gamma$, as calculated by use of Eq.\ref{eq:ah}, over time for constant background density (solid), weak gradient (dotted), strong gradient (dashed).}
\end{figure}
In order to quantify the acceleration efficiency, we calculate the amount of electrons that have energies higher than $E_{kin}^*=(15 v_{th})^2 m_e \gamma$. This number is calculated as a share of the total number of electrons in the system and followed over time, i.e.
\begin{equation}\label{eq:ah}
  AH = \frac{\int_{E_{kin}^*}^{E_{kin}^{max}} f(E_{kin}) \, \mathrm{d}E_{kin}}{\int_{0}^{E_{kin}^{max}} f(E_{kin}) \, \mathrm{d}E_{kin}} .
\end{equation}
Early stages of the simulation, when effects due to the non-zero initial current dominate, are disregarded. The corresponding graph is shown in Fig.\ref{fig:apf}. Clearly, the gradient has the effect of increasing the fraction of accelerated particles. While the weak gradient curve (dotted) tends to sit just on top of the constant density one (solid), the strong gradient case (dashed) lies much higher.
\begin{figure}
\includegraphics[scale=0.49]{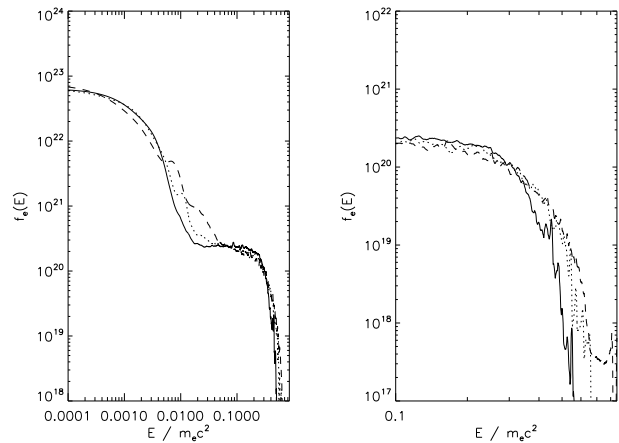}
\caption{\label{fig:apfcomp} Left: Final simulation snapshots of Figs.\ref{fig:especf0} (solid), \ref{fig:especf3} (dotted), \ref{fig:especf10} (dashed) plotted for comparison. Right: Zoomed in on high energy tail for clarity.}
\end{figure}
Fig.\ref{fig:apfcomp} shows that the gradient not only increases the share of accelerated particles, but accelerates particles to higher energies, respectively velocities.
\\
\begin{figure}
\includegraphics[scale=0.49]{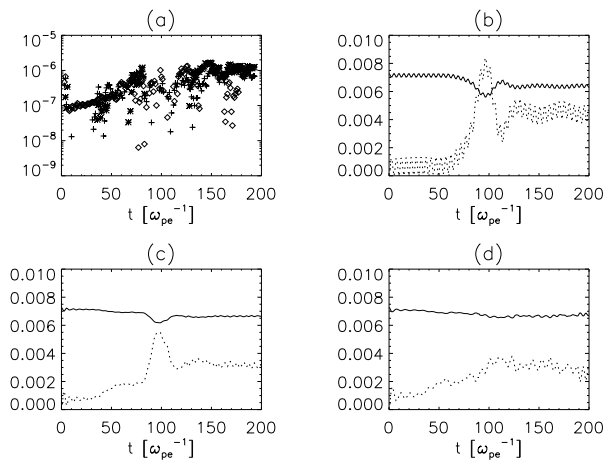}
\caption{\label{fig:EBfig} (a): $[E(t)-E(0)]/E(0)$ for constant background (diamond), weak gradient (cross), strong gradient (asterisk). (b): kinetic particle energy (solid) and Langmuir wave energy (dotted) for constant background. Langmuir wave energy multiplied by a factor $5$ for clarity. (c): as top right but for weak gradient. (d): as top right but for strong gradient.}
\end{figure}
Fig.\ref{fig:EBfig} shows the time evolution of respective energies for the various runs. Fig.\ref{fig:EBfig}a shows a total energy conservation check. It plots the quantity $[E(t)-E(0)]/E(0) = \Delta E/E(0)$, giving the change of total energy normalized to the initial total energy in the system as a function of time. We obtain similar results for all runs, with a maximum energy error of $\Delta E/E(0) \approx 10^{-6}$, which is a satisfactory value for the energy conservation. The panels (b), (c) and (d) of Fig.\ref{fig:EBfig} show the interplay between kinetic energies of particles and Langmuir wave energy. It should be noted, that for clarity of the plot, wave energies have been multiplied by a factor $5$. Fig.\ref{fig:EBfig} shows that, the particle energy is (partially) converted into wave energy and then transferred back to the particles. From Fig.\ref{fig:EBfig}b and Fig.\ref{fig:EBfig}c, it should be noted, that at around $\approx 100 \omega_{pe}^{-1}$ the Langmuir wave energy has a pronounced peak. This is the consequence of Langmuir wave growth via the beam plasma instability seen in Figs \ref{fig:tdf0} and \ref{fig:tdf3}. In the strong gradient case (Fig.\ref{fig:EBfig}d) Langmuir wave growth seems less vigorous (see Fig.\ref{fig:tdf10}). However, as can be seen in Fig.\ref{fig:f10loc} this is offset by the effect of drift in $k$-space towards lower wavenumbers $k$. Note that despite the fact that Langmuir waves are accelerating electrons, the net electron kinetic energy is decreasing. This is because the initial beam energy is converted to produce Langmuir waves. However, overall the drift in $k$-space produces a significant population of highly super-thermal electrons. The $k$-space drift will be discussed in section \ref{sec:locb}.
%
\section{\label{sec:locb} Localized Beam Injection}
%
In the previous section, it was established that the background density gradient has a clear effect on the fraction of accelerated electrons. It was also shown, that Langmuir waves were being generated via the beam-plasma instability. Ref.\cite{2012A&A...539A..43K} shows that in their study Langmuir waves would drift to smaller wavenumbers $k$, allowing them to increase their phase speed, $v_{ph}=\omega/k$, and, therefore, being subject to absorption by faster electrons. In the previous section, it was impossible to tell if such a drift was actually present. The reason being, that due to the non-localized nature of the beam, waves were being generated everywhere in the plasma. Hence, at every point along $x$ waves of different $k$ were {\it excited}, interfering with travelling waves from other sources in the plasma, which would {\it drift} to the wavenumber in question. In order to analyse if such a drift takes place, a localized beam was injected and left free to penetrate the plasma. A case with constant background density (as section \ref{sec:f0}) was considered along with the strong gradient case used in section \ref{sec:f10}. The beam peak density is given by the beam density in the previous sections. However, the beam has a finite width and its density profile is given by
\begin{equation}\label{eq:beamprof}
  n_b(x) = n_b \exp[-[(x-x_{max}/25)/(x_{max}/40)]^8] .
\end{equation}
\\
%
\subsection{\label{sec:f0loc} Constant Background Plasma}
%
%
\begin{figure*}
\includegraphics[scale=0.80]{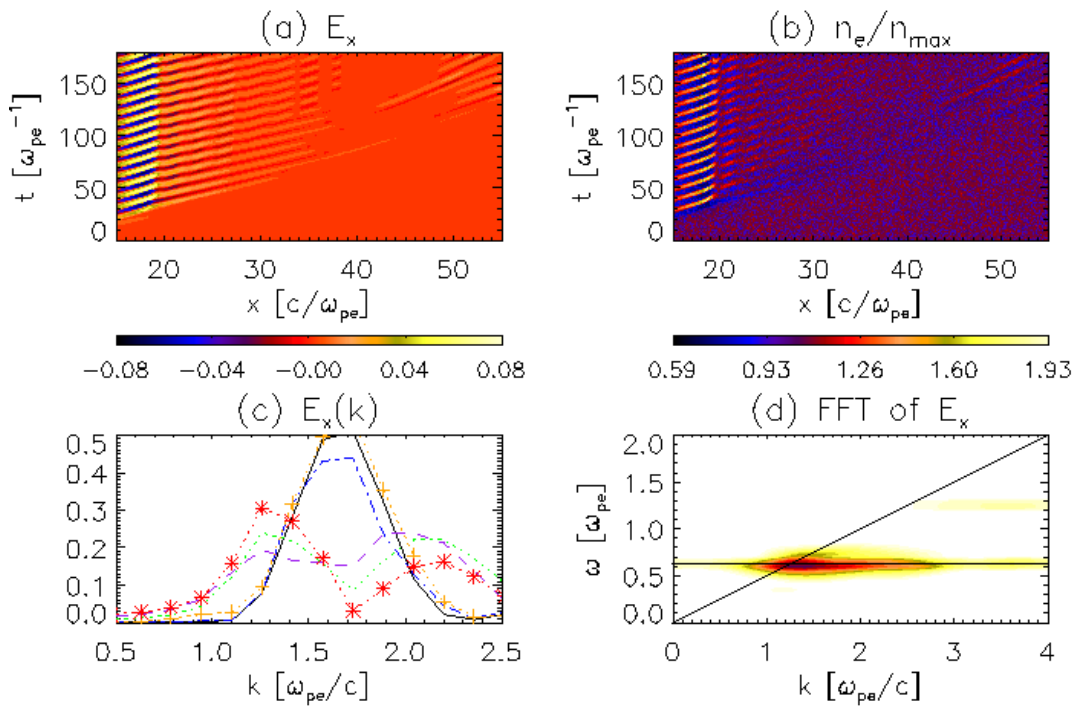}
\caption{\label{fig:f0loc} Localized beam injection with constant background density. (a): time-distance plot for $E_x$ component. (b): time-distance plot for background electron density. (c): $E_x(k)$ for $t=42 \omega_{pe}^{-1}$ (solid, black), $t=60 \omega_{pe}^{-1}$ (dashed, purple), $t=87 \omega_{pe}^{-1}$ (dash-dotted, blue), $t=105 \omega_{pe}^{-1}$ (dotted, green), $t=122 \omega_{pe}^{-1}$ (dotted with +, orange), $t=140 \omega_{pe}^{-1}$ (dotted with *, red). Y-axis in arbitrary units. (d): 2D Fourier transform of $E_x(x,t)$. Note that for clarity in panel (d), the color scheme is inverse to the ones in the upper row.}
\end{figure*}
Injection of a spatially localized beam generates a current at the beam injection region. Everywhere else in the box, the initial parameters correspond to a zero-current state. Therefore, in the beam injection region, a standing wave can be observed from the start of the simulation. This wave oscillates at the electron plasma frequency. It is undesirable to include information from this standing wave when performing a Fourier analysis, thus we consider only the region to the right of the beam injection, i.e. the region the beam (and waves) is (are) travelling into. We also cut off redundant regions on the right, where the waves never make it to during the course of the simulation. Fig.\ref{fig:f0loc} shows the results of the simulation. Fig.\ref{fig:f0loc}a shows the $E_x$ component. The plot shows clear waves that are being excited on the left edge (or just beyond) and travelling towards the right. It is already noticeable that the slopes of the waves change as they propagate. Fig.\ref{fig:f0loc}b shows the background density evolution, maintaining the correlation to the electric field via Gauss's law. Fig.\ref{fig:f0loc}c shows the time evolution of the spatial Fourier transform of $E_x$. In the case of constant background density, we expect wavenumber drifts to be a result of non-linear wave-wave interactions (see Ref.\cite{2012A&A...539A..43K} and references therein). We can see a clear tendency of wave power being shifted, as well as damped i.e. re-absorbed by plasma particles, leading to energy redistribution among electrons. It should be noted that, this is similar to the results in Ref.\cite{2012A&A...544A.148K}. Fig.\ref{fig:f0loc}d shows the full 2D Fourier transform of $E_x(t)$ (i.e. of Fig.\ref{fig:f0loc}a). The curves follow Langmuir wave dispersion relation and the beam resonance condition. The majority of waves are clearly being generated at the intersection of the curves.\\
Movie4 in Ref.\cite{mov:ref} shows the evolution of the densities for a localized beam injection into constant background plasma. Clearly the beam is dispersed as it propagates due to quasi-linear relaxation. At one occasion the beam density even overcomes that of the background, resulting in a strong signal in the top panels of Fig.\ref{fig:f0loc}. Soon the peak density of the beam is reduced and wave growth is no longer favourable, hence, Fig.\ref{fig:f0loc} shows propagating waves, but no new wave generation.
\\
%
\subsection{\label{sec:f10loc} Strong Gradient Case}
%
%
\begin{figure*}
\includegraphics[scale=0.80]{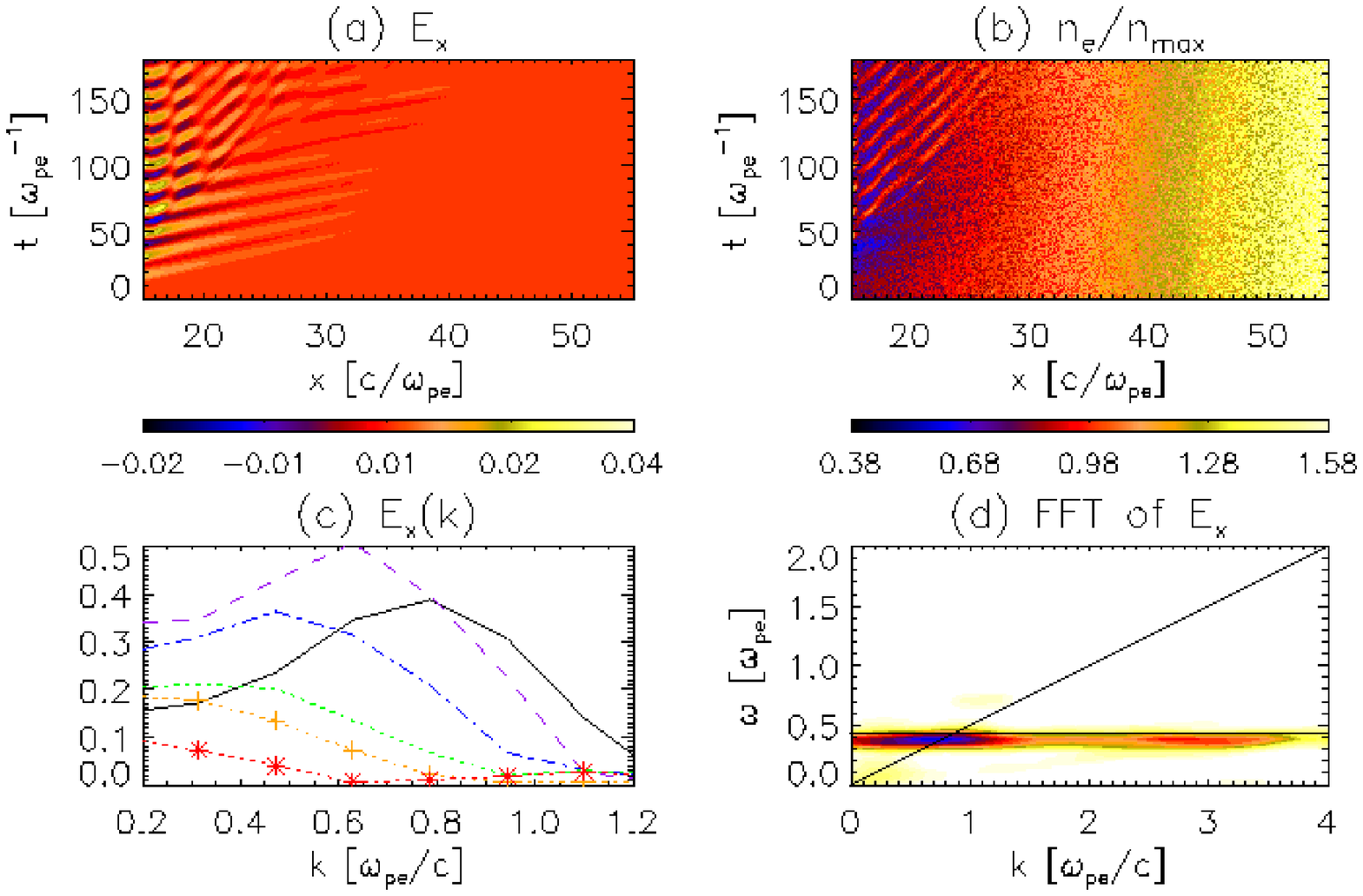}
\caption{\label{fig:f10loc} As Fig.\ref{fig:f0loc} but for the strong gradient case.}
\end{figure*}
The introduction of a background density gradient is expected to have a strong influence on the outcome of this study, as it amplifies the roles of wave dissipation and refraction. Fig.\ref{fig:f10loc} shows that - similar to the constant density case - waves are being generated on the left edge of the plot (or just beyond) and propagate towards the right. However, we can immediately see, that those waves are no longer allowed to propagate as far, due to the density gradient. The key feature of the plot is Fig.\ref{fig:f10loc}c, where one can clearly see, that the $k$-drift shows a clear shift to smaller wavenumbers over time. The wave power drifts from $\approx 0.6 \omega_{pe}/c$ to $\approx 0.4 \omega_{pe}/c$, where evident damping occurs. Also, in line with Fig.\ref{fig:f10loc}d, excited wavenumbers are smaller due to the shifted resonance condition in the region of interest. Movie5 of Ref.\cite{mov:ref} shows the evolution of the densities. Similar to the constant case, the beam is dispersed and its peak density is strongly decreased as it propagates through the plasma.\\
%
\section{\label{sec:conclusion} Conclusions}
%
The main goal of this study was to explore a mechanism that offers a potential solution to the problem of high intensity hard x-ray (HXR) emission observable during solar flares. This study seeks to extend previous work done in the field. A previous collisional quasi-linear theory study \cite{2012A&A...539A..43K} showed the importance of a background plasma density gradient, while a previous PIC analysis \cite{2012A&A...544A.148K} offered a fully kinetically self-consistent investigation. This study extends Ref.\cite{2012A&A...539A..43K} by the use of a self-consistent, fully kinetic approach and extends Ref.\cite{2012A&A...544A.148K} by introduction of a background plasma density gradient. The observed HXR spectra are thought to be evidence of a higher-than-usual population of high energy electrons, which are thought to be a result of Langmuir wave generation and absorption. This study sheds light on how a background density gradient influences the electron acceleration process in the fully kinetic regime. 3D fully relativistic, electromagnetic particle-in-cell (PIC) simulations with realistic mass ratio were performed. A mono-energetic beam of high energy electrons was injected into a magnetized, Maxwellian, homogeneous and inhomogeneous plasma. The initial electron distribution function in phase space has a bump in the forward direction, making the system unstable to the beam-plasma instability. Quasi-linear theory suggests that such a situation will allow Langmuir wave growth with subsequent plateau formation in the distribution function. Both effects were successfully demonstrated in our simulations. Waves were identified to be electrostatic by fulfilling Gauss's law. Generation of waves was shown to happen at the resonance of the dispersion relation for Langmuir waves and the beam mode. The main focus of the present study was to investigate the role of the background density gradient in the context of the acceleration of electrons. Three different cases with unbound beam injection were investigated: a) a constant background; b) a weak density gradient, $n_{e,R}/n_{e,L}=3$; c) a strong gradient case, $n_{e,R}/n_{e,L}=10$. It could be shown that the strong gradient case produced the largest fraction of electrons that have velocities above $15 v_{th}$. The weak gradient case also showed an increased number of high energy electrons. Further, two runs with localized beam injections were performed in order to analyse wave properties such as wavenumber drifts. The spatially localized beam was injected into both a constant background density profile, as well as a strong gradient one. The evolution of the wave power with respect to the wavenumber $k$ was analysed. It was shown that the Langmuir wave power indeed drifts to smaller wavenumbers, which is in line with a previous quasi-linear theory study \cite{2012A&A...539A..43K}. It should be noted, that computational constraints only allow us an investigation of a density rise of a factor $n_{e,R}/n_{e,L}=10$, whereas in a real situation, where electrons race down coronal loops from a flare region towards the chromosphere, the density increase would be a factor of $\approx 10^4$. The cumulative effect of this much larger density increase would be a much larger number of accelerated electrons, which could potentially account for the observed HXR radiation. Despite the fact that the main focus of this paper was the HXR emission in the solar flare context, all findings are also likely to be applicable to the Earth bow-shock \cite{1990JGR....95.4175O,1994JGR....9923481Y,2011AnGeo..29..613K}.\\
\\
{\bf ACKNOWLEDGEMENTS.}\\
The authors are financially supported by the HEFCE-funded South East Physics Network (SEPNET) UK. D.T.’s research is supported by The Leverhulme Trust Research Project Grant RPG-311 and STFC Grants ST/J001546/1 and ST/H008799/1.


\end{document}